\documentclass{kluwer}    % Specifies the document style.

\usepackage{graphicx}
\usepackage{amsthm}   % See h.c. printout for how to use this
\usepackage{amssymb}  % assumes amssymb package installed
\usepackage{mathrsfs} % for Lagrangian symbol
\usepackage{stmaryrd} 
\usepackage{txfonts} % for the varprod

\newtheorem{example}{Example}

\begin{document}                                                                                   
\begin{article}
\begin{opening}         
\title{What does Newcomb's Paradox Teach us?}
\runningtitle{Newcomb teaches us}
%\author{David H. Wolpert$^{1}$ \& Gregory Benford$^2$ \\ \\
% 1 -  NASA Ames Research Center,
% MS 269-1, Moffett Field, CA 94035-1000\\
% (650) 604-3362 (V), (650) 604-3594 (F), \texttt{david.h.wolpert.nasa.gov}\\
%2 - Physics and Astronomy Department, 
%University of California, \\ Irvine, CA 92692
%}
%%\title{A Sample Document\thanks{Footnote 
%%            to the title with the `thanks' command.}} 
%%\author{Leslie \surname{Lamport}}  
%%\runningauthor{Leslie Lamport}
%%\runningtitle{A Sample Document}
%%\institute{Author's Affiliation}
%\date{April 15, 1993}
\date{}

\begin{abstract}
In Newcomb's paradox you choose to receive either the contents of a
particular closed box, or the contents of both that closed box and
another one.  Before you choose though, an antagonist uses a
prediction algorithm to deduce your choice, and fills the two boxes
based on that deduction. Newcomb's paradox is that game theory's
expected utility and dominance principles appear to provide
conflicting recommendations for what you should choose. A recent extension of game theory
provides a powerful tool for resolving paradoxes concerning
human choice, which formulates such paradoxes in terms of Bayes nets. Here we apply this tool to Newcomb's scenario.
We show that the conflicting recommendations in Newcomb's
scenario use different Bayes nets to relate your
choice and the algorithm's prediction.  These two
Bayes nets are incompatible. This resolves the paradox: the reason
there appears to be two conflicting recommendations
is that the specification of the underlying Bayes net is 
open to two, conflicting interpretations. We then
show that the accuracy of the
prediction algorithm in Newcomb's paradox, the focus of much previous
work, is irrelevant.  We similarly show that the utility functions of you
and the antagonist are irrelevant. We end by showing that Newcomb's paradox is
time-reversal invariant; both the paradox and its resolution are
unchanged if the algorithm makes its `prediction'
\emph{after} you make your choice rather than before.
\end{abstract}

%\newpage
%\noindent \textbf{Sections:} 

%$1\;\;\;\;$Introduction

%$\;\;\;\;\;\;1.1\;\;\;\;$Background

%$\;\;\;\;\;\;1.2\;\;\;\;$Game theory over Bayes nets

%$2\;\;\;\;$Resolving Newcomb's paradox by extending game theory 

%$\;\;\;\;\;\;2.1\;\;\;\;$The first decomposition of the joint probability

%$\;\;\;\;\;\;2.2\;\;\;\;$The second decomposition of the joint probability

%$\;\;\;\;\;\;2.3\;\;\;\;$Combining the decompositions


\end{opening}

\newpage

\section{Introduction}
\subsection{Background}
Suppose you meet a Wise being ($W$) who tells you it has put
\$1,000 in box A, and either \$1
million or nothing in box B. This being tells you to either take the
contents of box B only, or to take the contents of both A and B.
Suppose further that the being had put the \$1 million in box B 
if a prediction algorithm used by the being had said that you
would take only B. If instead the algorithm had predicted you would take both
boxes, then $W$ put nothing in box B.

Presume that due to determinism, there exists a perfectly accurate
prediction algorithm, and assume that it is this perfect prediction
algorithm that $W$ uses. Suppose further that when you choose which
boxes to take, you don't know the prediction of that algorithm. What
should your choice be?

In Table 1 we present this question as a game theory
matrix involving $W$'s prediction and your choice. Two seemingly logical
answers to your question contradict each other. The Realist answer is that you should
take both boxes, because your choice occurs
after $W$ has already made its prediction, and since you have free will, you're
free to make whatever choice you want, independent of that prediction that
$W$ made. More precisely, if $W$
predicted you would take A along with B, then taking both gives you
\$1,000 rather than nothing. If instead $W$ predicted you would take only B,
then taking both boxes yields \$1,001,000, which again is \$1000
better than taking only B. 

In contrast, the Fearful answer is that $W$ designed a prediction
algorithm whose answer will match what you do. So you can get
\$1,000 by taking both boxes or get \$1
million by taking only box B.
Therefore you should take only B.

This is the conventional formulation of Newcomb's Paradox, a famous
logical riddle stated by William Newcomb in
1960~\cite{nozi69,gard74,bama72, jaco93,huri78,cala85,levi82,gean97,coll01,pisl02b,burg04}.  Newcomb never
published the paradox, but had long conversations about it with with
philosophers and physicists such as Robert Nozick and Martin Kruskal,
along with Scientific American's Martin Gardner. Gardner said after
his second Scientific American column on Newcomb's paradox appeared
that it generated more mail than any other column.

One of us (Benford) worked with Newcomb, publishing several papers
together. We often discussed the paradox, which Newcomb invented to
test his own ideas. Newcomb said that he would just take B; why fight
a God-like being? However Nozick said, ``To almost everyone, it is
perfectly clear and obvious what should be done. The difficulty is
that these people seem to divide almost evenly on the problem, with
large numbers thinking that the opposing half is just being
silly''~\cite{nozi69}.
	
It was Nozick who pointed out that two accepted principles of game
theory appear to conflict in Newcomb's problem. The expected-utility
principle, considering the probability of each outcome, says you
should take box B only. But the dominance principle argues that if one
strategy is always better than the other strategies no matter what
other players do, then you should pick that strategy~\cite{futi91,
myer91, osru94}.) No matter what box B contains, you are \$1000 richer
if you take both boxes than if you take B only. So the dominance
principle says you should take both boxes.

Is there really a contradiction? Some philosophers argue that a
perfect predictor implies a time machine, since with such a machine
causality is reversed, i.e., for
predictions made in the present to be perfectly determined by events in the future
means that the future causes past events.{\footnote{Interestingly, near when Newcomb devised
the paradox, he also coauthored a paper proving that a tachyonic time
machine could not be reinterpreted in a way that precludes such
paradoxes~\cite{bebo70}. The issues of time travel and Newcomb-style paradoxes are
intertwined.}}  
	
But Nozick stated the problem specifically to exclude backward
causation (and so time travel), because his formulation demands only
that the predictions be of high accuracy, not perfect. So arguments
about time travel cannot resolve the issue. 
%Whether or not one believes time travel is involved, faced with Newcomb's 
%seemingly logical paradox, it would seem that the
%conclusion must be that perfect prediction is impossible.
%Worse still, Nozick's
%reformulation seems to imply that the (in)fallibility of $W$'s
%prediction provides yet another conundrum, in addition to the one
%underlying Newcomb's paradox.

Recently an extension to game theory has been introduced that provides
a powerful tool for resolving apparent logical paradoxes concerning human choice.
This tool formulates games in terms of Bayes nets.
Here we illustrate this tool by
using it to resolve Newcomb's paradox. In a nutshell,
it turns out that Newcomb's scenario does not fully specify the Bayes net underlying the
game you and $W$ are playing. The two ``conflicting principles of game theory" actually
correspond to two different underlying Bayes nets. As we show, those nets are mutually
inconsistent. So there is no conflict of game theory principles --- simply
imprecision in specifying the Bayes net underlying the game you and $W$ and playing.

The full extension of game theory is more powerful than needed to
resolve Newcomb's paradox; a smaller extension would suffice.  However
using the full extension to analyze a paradox often draws attention to
aspects of the paradox that otherwise would escape notice. That is the
case here. We use the full extension to show that the accuracy of the
prediction algorithm in Newcomb's paradox, the focus of much previous
work, is irrelevant.  In addition we show that the utility functions
of you and $W$ are actually irrelevant; the self-contradiction built
in to Newcomb's scenario has to do with imprecision in specification
of the underlying Bayes net, and is independent of the utility
functions. We also show that Newcomb's paradox is time-reversal
invariant; both the paradox and its resolution are unchanged if the
algorithm makes its `prediction'
\emph{after} you make your choice rather than before.

In the next section we present a simplified version of our full
argument. We then review the recent extension of game theory.  In the
following section we use extended game theory to provide our fully
detailed resolution of Newcomb's paradox. We end with a discussion.

\section{Decompositions of the reasoning of Fearful and Realist}

There are two players in Newcomb's paradox: you, and the wise being
$W$. In addition, there are two game variables that are central to the
paradox: $W$'s prediction, $g$, and the choice you actually make,
$y$. So the player strategies will involve the joint probability
distribution relating those two variables. Since there are only two
variables, there are only two ways to decompose that joint probability
distribution into a Bayes net. These two decompositions turn out to
correspond to the two recommendations for how to answer Newcomb's
question, one matching the reasoning of Realist and one matching
Fearful.

Define $ab$ as the event that $W$ predicts you will take both boxes,
and $b$ as the event that $W$ predicts you will only take box
$B$. Similarly define $AB$ as the event that you actually take both
boxes, and $B$ as the event that you actually take only box $B$. So
$P(g = ab \mid y = AB) = P(ab \mid AB)$ is the probability that $W$
predicts correctly, given that you choose $AB$. Similarly $P(b \mid
B)$ is the probability that $W$ predicts correctly given that you
choose only $B$.  

Decision theory says that you should choose $P(y)$ so
as maximize the associated expected utility.
To make that formal, first we express Fearful's
reasoning as a joint probability distribution over $g$ and $y$:
$$
P(y,g)=P(g\mid y)P(y)
$$
where $P(g \mid y)$ is pre-fixed, by $W$'s
prediction algorithm.

Decision theory says that for this decomposition, you should set
$P(AB)$ (and therefore $P(B) = 1 - P(AB)$) so as to maximize
\begin{eqnarray*}
1000 [P(ab \mid AB)P(AB)] &+& 1001000 [P(b \mid AB)P(AB)] \\
&+& 0[P(ab \mid
B)P(B)] \;+\; 1000000[P(b \mid B)P(B)].
\end{eqnarray*}
Provided that the associated distributions
$P(ab \mid AB)$ and $P(b \mid B)$ are large enough ---
provided $W$'s prediction algorithm is accurate enough ---  to achieve
this maximization you should set $P(B) = 1$. In other words,
you should choose to take only $B$. This expresses 
Fearful's ``expected utility'' reasoning.

On the other hand, under Realist's reasoning, you can make your
decision entirely independently of $W$'s prediction. This amounts
to the assumption that 
\begin{eqnarray*}
       P(y,g)&=&P(y \mid g)P(g) \\
	&=& P(y) P(g)
\end{eqnarray*}
where you set $P(y)$ and $W$ sets $P(g)$.
For this decomposition, decision theory says that you should choose
$P(AB)$ to maximize
\begin{eqnarray*}
1000 [P(ab)P(AB)] &+& 1001000 [P(b)P(AB)] \\
&+& 0[P(ab)P(B)] \;+\; 1000000[P(b)P(B)].
\end{eqnarray*}
which is achieved by setting $P(B) = 0$, no matter what $P(g)$ is.
This conclusion expresses the dominance principle of game theory.

This foregoing analysis illustrates how the reasoning of Fearful and
Realist correspond to different decompositions of the joint
probability of your choice and $W$'s choice. The simple fact that
those two decompositions differ is what underlies the resolution of
the paradox --- you can state Newcomb's scenario in terms of Fearful's
reasoning, or in terms of Realist's reasoning, but not
both.{\footnote{We are grateful to an anonymous referee for
emphasizing the underlying simplicity of our analyzing Newcomb's scenario
by decomposing the reasoning of Fearful and Realist into different
joint probability distributions.}}

The foregoing analysis, while quite reasonable, does not formally
establish that there is no other way to interpret Newcomb's scenario
besides the two considered.  For example, one might worry that there
is a way to merge the joint probability decompositions of Fearful and
Realist, to get a hybrid decomposition that would somehow better
captures Newcomb's scenario. In addition, it is not clear if Newcomb's
scenario should be viewed as a decision theory problem, in which case
$W$'s behavior is pre-fixed (as assumed above). Much of the literature
instead casts Newcomb's paradox as a game theory problem, to allow $W$
to act as an antagonist (e.g., by giving $W$ a utility function that
equals the negative of your utility function). In particular, viewing
Newcomb's scenario as a game raises questions like how the analysis
changes if $W$ can choose the accuracy of the prediction algorithm,
rather than have it be pre-fixed. Finally, the simplified analysis
presented above does not draw attention to subtleties of Newcomb's
scenario like its time-invariance.

To overcome these shortcomings it is necessary to use extended game
theory. That is the topic of the next two sections.

\section{Extended game theory}

Central to Newcomb's scenario is a prediction process, and its
(in)fallibility.  
Recent work 
%has revealed unavoidable limitations on
%the accuracy of \emph{any} prediction process. Amongst other things, this work
proves that any given prediction algorithm must fail on at least one
prediction task~\cite{bind08,wolp08b,wolp10}. That particular result does not
resolve Newcomb's paradox. However the proof of that result requires
an extension of conventional game theory.  And as we show below, that
extension of game theory provides a way to resolve Newcomb's paradox.

\subsection{Extended games}

To introduce extended game theory, consider a situation where
there is a set of several players, $\mathscr{N}$, together with a set of \emph{game variables}. 
A \emph{(profile) strategy} is a set of one or more joint probability distributions over the game variables.
Every player  has their own set of strategies (i.e., their own set of
sets of distributions) that they can choose from. In general we indicate
a member of player $i$'s strategy set by $S_i$. 

If the intersection of every possible joint choice of strategies by
all the players is a single joint probability distribution, we say the game is \emph{proper}. 
For a proper game, it doesn't matter what choice each of the players in $\mathscr{N}$ makes;
their joint choice always specifies a single joint probability distribution over the game variables.
All the games considered in conventional game theory are proper.
For the rest of this subsection, we restrict attention to proper games.

Every player in an extended game whose profile strategy
set contains exactly one strategy (i.e., one set of probability
distributions)  is called a \emph{Nature} player. Note that the strategy ``chosen" by a
Nature player is automatic. 
%However to get a joint strategy choice by all the players,
%we must specify a mechanism for each non-Nature player $i$
%to choose one of the (multiple) strategies in their strategy set.

%To that end, e
Every non-Nature player $i$ has their own \emph{utility function} $u_i$
which maps any joint value of
the game variables to a real number. So given any joint choice of
strategies, in a proper game the associated (unique) joint distribution provides the
expected utility values of the non-Nature players. 
%As described below, this forms
%the basis for each non-Nature player to choose a strategy.

To play the game, all the non-Nature
players independently choose a strategy (i.e., choose a set of
joint distributions over the game varaibles) from their respective strategy sets.
Their doing this fixes the joint strategy, and therefore the joint distribution
over the game variables (if the game is proper), and therefore the expected utilities of all the non-Nature players. 

Game theory is the analysis of what strategies the non-Nature
players will jointly choose in this situation, under different possible
choice-making principles. One such principle is the dominance
principle, mentioned above, and another is the expected-utility
principle, also mentioned above.

The following example shows how to formulate a game very popular in  conventional
game theory in terms of extended game theory:

\begin{example} Matching pennies.
\begin{enumerate}
\item In the conventional formulation of the  ``Matching Pennies" game, there are two players, Row and
Col, both of whom are non-Nature players. There are also two game variables, $X_{R}$ and $X_{C}$, with
associated values $x_R, x_C$, both of which can be either $0$ or $1$. To play the
game player Row chooses a probability distribution $p_R(x_R)$ and
player Col independently chooses a probability distribution
$p_C(x_C)$.  Taken together, those two distributions define the joint
distribution over the game variables according to the rule $P(x_R,
x_C) = P(x_R) P(x_C)$.

Player Row's utility for any joint value ($x_R, x_C)$ formed by
sampling $P(x_R, x_C)$ equals $1$ if $x_R = x_C$, $0$
otherwise. Conversely, Player Col's utility is $0$ if $x_R = x_C$, $1$
otherwise. So the expected utility of player Row under $P(x_R, x_C)$
is 
$$
\sum_{x_R, x_C} p_R(x_R)p_C(x_C) \delta_{x_R, x_C},
$$
where the Kronecker delta function $\delta_{a, b}$ equals $1$ if $a = b$, and $0$ otherwise.
Similarly, the expected utility of player Col is 
$$
\sum_{x_R, x_C} p_R(x_R)p_C(x_C) [1 - \delta_{x_R, x_C}].
$$

\item We now show how to reformulate Matching Pennies as an extended game. 
Each of player Row's strategies is specified by a
single real number, $p_R$. The strategy $S_{p_R}$ associated
with the value $p_R$ is the set of all joint distributions $Pr(x_R,
x_C)$ obeying $Pr(x_R, x_C) = [p_R \delta_{x_R, 1} + (1 - p_R)
\delta_{x_R, 0}]Pr(x_C)]$ for some distribution $Pr(x_C)$.  In other
words, it is the set of all joint distributions $P(x_R, x_C)$ whose marginal $P(x_R)$
is given by $p_R$. Similarly,
each of player Col's strategies $S_{p_C}$ is specified by a
single real number, $p_C$. The strategy $S_{p_C}$ associated
with the value $p_C$ is the set of all joint distributions $Pr(x_R,
x_C)$ such that $Pr(x_R, x_C) = Pr(x_R)[p_C
\delta_{x_C, 1} + (1 - p_C) \delta_{x_C, 0}]$ for some distribution
$Pr(x_R)$.

In this extended game formulation, to play the game, each player
independently chooses a strategy. Whatever strategy
$S_{p_R}$ Player Row chooses, and whatever strategy $S_{p_C}$ Player
Col chooses, there is a unique joint distribution consistent with
their choices, 
\begin{eqnarray*}
Pr(x_R, x_C) &=& S_{p_R} \cap S_{p_C} \\
&=& [p_R \delta_{x_R, 1} + (1 -
p_R) \delta_{x_R, 0}] [p_C \delta_{x_C, 1} + (1 -
p_C) \delta_{x_C, 0}]
\end{eqnarray*}
The expected
utilities of the two players are then given by their expected utilities
under $S_{p_R} \cap S_{p_C}$.
\end{enumerate}

$ $

\noindent A moment's thought shows that the extended game formulation 
is identical to the original formulation in (1).
\end{example}

\subsection{Over-played and under-played games}
\label{sec:improper}

If a game is
not proper, then there is a joint strategy, $S = \{S_i : i \in \mathscr{N}\}$, such that 
one of two conditions holds: 
\begin{enumerate}
\item $\cap_{i \in \mathscr{N}} S_i$ is empty. In this case we say the game is \emph{over-played}.
\item  $\cap_{i \in \mathscr{N}} S_i$ contains more than one distribution
over the game variables. In this case we say the game is \emph{under-played}.
\end{enumerate}

Note that one way to establish 
that $\cap_{i \in \mathscr{N}} S_i = \varnothing$ is to show that the sets
of conditional distributions $\{S_i\}$, taken together, violate the laws of probability.
(This is how we will do it below.)  In over-played games, the players are 
not actually independent; there is a joint strategy choice $S$ by the players that is impossible,
and therefore it is \emph{not} the case that every player $i$ is free to choose the associated
strategy $S_i$ without regard for the choices of the players.
Accordingly, if one specifies a game theoretic scenario where 
$\cap_{i \in \mathscr{N}} S_i = \varnothing$ for some joint strategy choice $S$, one
cannot at the same time presume that the players are independent. If you do so, then your specification
of the game theoretic scenario contradicts itself. 

Conversely, in under-played games, there is a joint strategy choice by the players
that does not uniquely determine the distribution over the game variables.
Accordingly,  if one specifies a game theoretic scenario where 
$\cap_{i \in \mathscr{N}} S_i$ contains more than one element for some joint strategy choice $S$,
then the outcome of the game is not well-defined. 

Often a ``paradox" involving human choice is 
a game theoretic scenario that is over-played, and therefore has a self-contradiction built into it,
or is under-played,
and therefore not well-defined. To resolve either type of paradox one simply requires that the
game theoretic scenario be proper. In particular, this is how Newcomb's paradox
is resolved.

\subsection{Representing extended game theory in terms of Bayes nets}

A \emph{Bayes net} over a set of variables $\{A, B, C,
\ldots\}$ is a compact, graphical representation of a joint probability distribution over
those variables~\cite{komi03,pear00}. Often in extended game theory it is convenient to express a player's 
profile strategy set in terms of a Bayes net. In particular, expressing strategy sets this
way provides a convenient way of establishing that a game is (not) proper,
and thereby analyzing apparent paradoxes involving human choice.
%The graph of the Bayes net specifies the
%conditional independencies of the joint distribution $P(A, B, C,
%\ldots)$. The Bayes net goes on to specify a conditional distribution
%relating the variables of each of those conditional independencies. In
%this way the Bayes net uniquely specifies the joint distribution.

Formally, any Bayes net contains three parts. The first part is a
Directed Acyclic Graph (DAG) comprising a set of vertices / nodes, $V$,
and an associated set of directed edges, $E \subseteq V \times V$. There are the same number of nodes 
in the DAG as there
are variables $A, B, C, \ldots$. A node $\nu'$ in the DAG such that  an edge
leads directly from $\nu'$ into a node $\nu$ is called a \emph{parent} of
$\nu$. The set of all parents of $\nu$ is written as pa$(\nu)$. Note that pa$(\nu)$
is the empty set for a root node.

The second part of a Bayes net is a one-to-one onto
function $\chi$ from the set of nodes of the DAG to the set of
variables $A, B, C,
\ldots$. So $\chi$ labels each of the nodes of the DAG with one
and only one of the variables $A, B, C, \ldots$. The final part of the
Bayes net is a set of conditional distributions, one for each node
$\nu \in V$ of the DAG. The conditional distribution at $\nu$ is a specification of
$P(\chi(\nu) \mid \chi[{\rm{pa}}(\nu)])$, where $\chi[{\rm{pa}}(\nu)]$
is the variables associated with those parent nodes. 

It is straight-forward to
see that any set of all such conditional distributions in the Bayes net uniquely 
specifies a joint distribution:
\begin{eqnarray}
P(A, B, C, \ldots) &=& \prod_{\nu \in V} P(\chi(\nu) \mid \chi[{\rm{pa}}(\nu)])
\label{eq:b_net}
\end{eqnarray}
Conversely, any joint distribution can be expressed as a Bayes net. (Indeed, in
general any joint distribution can be expressed as more than one Bayes net.)

Many games can be very naturally defined in terms of a Bayes net. Let
$(V, E)$ be the nodes and edges of the Bayes net. To define the game,
the nodes $V$ are partitioned among the players. So player $i$ is
assigned some subset $V_i \subseteq V$, for all $i, j \ne i$,
$V_i \cap V_j = \varnothing$
and $\cup_i V_i = V$. A profile strategy of player $i$ is simply
the specification $p^{V_i}$ of all the conditional distributions at
the nodes in $V_i$. More precisely, given a $p^{V_i}$, the
associated profile strategy is all joint distributions represented by
a Bayes net with DAG $(V, E)$ that have the
conditional distributions $p^{V_i}$ at the nodes in $V_i$, where the remaining
conditional distributions are unspecified. 

We are guaranteed
that any game defined in terms of a Bayes net this way is proper. In
other words, \emph{any} joint profile strategy of the players
corresponds to one and only one joint distribution over the game
variables. This is because a joint strategy fixes the conditional distributions
at all nodes in $V_i$ for all $i$ --- which means it fixes all the conditional
distributions. So a joint strategy uniquely specifies the full Bayes net.
Eq.~\ref{eq:b_net} then maps that Bayes net to the (unique) joint distribution over the game variables.

\begin{example}
In the Matching Pennies scenario, there are two variables, $x_R$ and
$x_C$. Therefore $V$ consists of two nodes, $\nu_R$ and $\nu_C$,
corresponding to $x_R$ and $x_C$ respectively. In addition, the two
variables are statistically independent, i.e., $P(x_R, x_C) =
P(x_R)P(x_C)$.  Therefore in the DAG there are no edges connecting the
two nodes.

A strategy of player Row is the specification of the conditional
distribution $P(\chi(\nu_R) \mid \chi[{\rm{pa}}(\nu_R)])$. Since
$\nu_R$ has no parents, this is just the distribution
$P(\chi(\nu_R))$. Similarly, a strategy of player Col is just the
distribution $P(\chi(\nu_C))$. Given the DAG of the game, 
plugging into Eq.~\ref{eq:b_net} shows that a joint
distribution is determined from a joint strategy by 
$P(\chi(\nu_R), \chi(\nu_C)) = P(\chi(\nu_R)) P(\chi(\nu_C))$.

In contrast, say that Row player had moved first, and there was a
noisy sensor making an observation $D$ of that move, and that Col chose
their move based on the value of $D$. Then the Bayes net would have
three nodes, $\nu_R, \nu_D$ and $\nu_C$, corresponding to the three
variables. The node $\nu_R$ would have no parents, the node $\nu_D$
would have (only) $\nu_R$ as its parent, and the node $\nu_C$ would
have (only) $D$ as its parent. We can have Row and Col be non-Nature players,
and  the sensor be a Nature player. So the Row player would set
$$
P(\chi(\nu_R) \mid \chi[{\rm{pa}}(\nu_R)]) = P(\chi(\nu_R)),
$$
by choosing a distribution from its strategy set. In addition
a Nature player would set 
$$
P(\chi(\nu_D) \mid \chi[{\rm{pa}}(\nu_D)]) =
P(\chi(\nu_D) \mid \chi(\nu_R)),
$$
to the single strategy in its strategy set. Finally, the Col player would set
$$
P(\chi(\nu_C) \mid \chi[{\rm{pa}}(\nu_C)]) =
P(\chi(\nu_C) \mid \chi(\nu_D)).
$$
by choosing a distribution from its strategy set.
\end{example}

\subsection{Conflicting DAG's}

All of the games arising in conventional game theory are formulated as a Bayes net game involving
a single underlying DAG. Accordingly, all those games are proper.

Often in a game theory paradox it is implicitly presumed that
the game is proper, just like conventional games. In
particular, it is often presumed that the players are free to choose their strategies independently of one another.
However in contrast to the case with games in conventional game theory, often in these paradoxes
the strategy sets of the different players are defined in terms of different Bayes net DAGs. Due to
this, the presumption of player independence might be violated.
Showing that this is the case --- showing 
%If we can show that in such a game
%
%Showing that the game in an apparent paradox
%is not proper is a powerful way to resolve such paradoxes.
%In particular, say we can show that a given apparent
%paradox implicitly defines the strategy sets of some of the players in terms of different
%Bayes net DAGs, rather than having all
%strategy sets defined in terms of the same DAG.  Since different DAGs are used for different players,
%the game may not be proper. We are not guaranteed that for any joint choice of strategies by the
%players, there is a joint distribution over the game variables 
%that is consistent with that joint choice. 
%The final step in resolving the given paradox is to show precisely this, that 
there are joint choices of conditional distributions by the players that are not simultaneously consistent
with any joint distribution --- shows that the implicit premise
in the paradox that the players are independent is impossible. Resolving such an apparent paradox simply means
requiring that the associated game be stated in
terms of a single Bayes net DAG, so that it is proper, and that therefore all
premises actually hold.

% Formally, in this extension
% of conventional game theory, multiple different Bayes nets relating
% the game variables are specified, and each player resets conditional
% distributions at some nodes in each of those Bayes nets however it desires. So a player's
% strategy now specifies conditional distributions in all of the Bayes
% nets at once. (In contrast, conventional game theory implicitly only
% allows a single Bayes net.)  This means that some joint player
% strategies may be impossible, in that they result in each Bayes net
% having a different joint distribution over the game
% variables. Whenever the joint distributions of two Bayes nets are
% inconsistent this way, some set of the conditional distributions in 
% one of the two nets will conflict with some set of distributions in
% the other Bayes net.

It is exactly such reasoning that underlies the
fallibility of prediction proven in~\cite{bind08,wolp08b,wolp10}. As we show
below, this reasoning also resolves Newcomb's paradox. Loosely
speaking, Newcomb's scenario involves two Bayes nets, one
corresponding to Realist's reasoning, and one corresponding to
Fearful's reasoning. One can choose one Bayes net or the other, but to
avoid inconsistency, one cannot choose both.  In short, Realist and
Fearful implicitly formalize Newcomb's scenario in different, and
incompatible ways.

In the next section, we first show how to express the reasoning of
Realist and of Fearful in terms of strategy spaces over two different
Bayes net DAGs.  We then show that those two Bayes net DAGs and associated
strategy spaces cannot be combined without raising the possibility of
inconsistent joint distributions between the Bayes nets.  This
resolves Newcomb's paradox, without relying on mathematizations of
concepts that are vague and/or controversial (e.g., `free will',
'causality'). In addition, our resolution shows that there is no
conflict in Newcomb's scenario between the dominance principle of game
theory and the expected utility principle. Once one takes care to
specify the underlying Bayes net game, those principles are completely
consistent with one another.

We end by discussing the implications of our analysis. In particular,
our analysis formally establishes the (un)importance of $W$'s
predictive accuracy. It also shows that the utility functions of you and $W$
are irrelevant. Furthermore, it provides an illustrative way to look at
Newcomb's scenario by running that scenario backwards in time.

%In this general approach, to analyze a prediction scenario one
%considers the joint probability distribution of only those random
%variables that occur in {\it{all}} possible formulations of the
%scenario. All other variables are treated as `hidden variables''
%underlying that single joint probability distribution. By doing that
%for Newcomb's scenario we can avoid choosing among its possible
%formulations.

\section{Resolving Newcomb's paradox with extended game theory}

There are two players in Newcomb's paradox: you, and the wise being
$W$. In addition, there are two game variables that are central to the
paradox: $W$'s prediction, $g$, and the choice you actually make,
$y$. So the player strategies will involve the joint probability
distribution relating those two variables. Since there are only two
variables, there are only two ways to decompose that joint probability
distribution into a Bayes net. These two decompositions turn out to
correspond to the two recommendations for how to answer Newcomb's
question, one matching the reasoning of Realist and one matching
Fearful.

\subsection{The first decomposition of the joint probability}

The first way to decompose the joint probability is
\begin{eqnarray}
P({y}, {g}) &=& P({g} \mid {y})  P({y}) 
\label{eq:1}
\end{eqnarray}
(where we define the right-hand side to equal 0 for any ${y}$ such
that $P({y}) = 0$). Such a decomposition is known as a `Bayes net'
having two `nodes'~\cite{pear00,komi03}. The variable $y$, and the
associated unconditioned distribution, $P(y)$, is identified with the
first, `parent' node. The variable $g$, and the associated conditional
distribution, $P(g \mid y)$, is identified with the second, `child'
node.  The stochastic dependence of $g$ on $y$ can be graphically
illustrated by having a directed edge go from a node labeled $y$ to a
node labeled $g$.

This Bayes net can be used to express Fearful's reasoning.  Fearful
interprets the statement that `W designed a perfectly accurate
prediction algorithm' to imply that $W$ has the power to set the
conditional distribution in the child node of the Bayes net, $P({g}
\mid {y})$, to anything it wants (for all ${y}$ such that $P({y})
\ne 0$). More precisely, since the algorithm is `perfectly
accurate', Fearful presumes that $W$ chooses to set $P({g} \mid {y}) =
\delta_{{g}, {y}}$, the Kronecker delta function that equals $1$ if ${g} = {y}$,
zero otherwise. So Fearful presumes that there is nothing you can do
that can affect the values of $P({g} \mid {y})$ (for all ${y}$ such
that $P({y}) \ne 0$). Instead, you get to choose the unconditioned
distribution in the parent node of the Bayes net, $P(y)$. Intuitively,
this choice constitutes your `free will'.

%Fearful makes one
%presumption for each of
%the two distributions on the right-hand side of Eq.~\ref{eq:1}. The
%first of the presumptions concerns you: Fearful presumes that your having
%`free will'' means that you have power to set $P({y})$ to anything  you want,
%with no constraints arising from what $W$ does.
%The second presumption in
%Fearful's reasoning concerns $W$. It says that $W$ has power
%to set the distribution $P({g} \mid {y})$ (for all ${y}$ such that $P({y})
%\ne 0$) to anything he wants, and that there is nothing you can do to affect this distribution. This presumption
%then says that what $W$ sets $P({g} \mid {y})$ to is the distribution
%$\delta_{{g}, {y}}$, the distribution that equals $1$ if ${g} = {y}$, zero
%otherwise. In other words, Fearful's second presumption is that for all ${y}$ such that $P({y}) \ne 0$, $W$'s prediction
%accuracy $P({g} \mid {y})$ is `pre-fixed'' to perfection, and that nothing you do
%can affect this. 

Fearful's interpretation of Newcomb's paradox specifies what aspect of
$P({y}, g)$ you can choose, and what aspect is instead chosen by
W. Those choices --- $P(y)$ and $P(g \mid y)$, respectively --- are
the `strategies' that you and $W$ choose. It is important to note that these
strategies by you and $W$ do \emph{not} directly specify the two variables
$y$ and ${g}$. Rather the strategies you and $W$ choose specify two different
distributions which, taken together, specify the full joint
distribution over $y$ and ${g}$~\cite{komi03}. This kind of strategy
contrasts with the kind considered in decision theory~\cite{berg85} or
causal nets~\cite{pear00}, where the strategies are direct specifications
of the variables (which here are $g$ and $y$).

In both game theory and decision theory, by the axioms of VonNeumann Morgenstern utility
theory, your task is to choose the strategy that maximizes your
expected payoff under the associated joint distribution.{\footnote{For
simplicity, we make the common assumption that utility is a linear
function of payoff~\cite{star00}. So maximizing expected utility is
the same as maximizing expected payoff.}}  For Fearful, this means choosing the $P({y})$
that maximizes your expected payoff under the $P({y}, {g})$ associated
with that choice.  Given Fearful's presumption that the Bayes net of
Eq.~\ref{eq:1} underlies the game with $P(g \mid y) = \delta_{g,y}$,
and that your strategy is to set the distribution at the first node,
for you to maximize expected payoff your strategy should be $P({y}) =
\delta_{{y}, B}$. In other words, you should choose $B$ with probability
$1$. Your doing so results in the joint distribution $P({y}, {g})
\;=\; \delta_{{g}, {y}} \delta_{{y}, B} \nonumber \; =\;
\delta_{{g}, B}  \delta_{{y}, B}$,
with payoff $1,000,000$. This is the formal justification of Fearful's
recommendation.{\footnote{In Fearful's Bayes net, $y$ `causally
influences $g$', to use the language of causal nets~\cite{pear00}. To
cast this kind of causal relationship in terms of conventional game
theory, we would have to replace the single-stage game in Table 1 with
a two-stage game in which you first set $y$, and
\emph{then} $W$ sets their strategy, having observed $y$.  This two-stage game is
incompatible with Newcomb's stipulation that $W$ sets their strategy before you do,
not after. This is one of the reasons why it is necessary to use
extended game theory rather than conventional game theory to formalize
Fearful's reasoning.}}

Note that since this analysis treats the strategy set of $W$
as being a single set of joint distributions, it can be interpreted as either a
decision-theoretic scenario or a game-theoretic one, with $W$ being a Nature player.

\subsection{The second decomposition of the joint probability}

The second way to decompose the joint probability is
\begin{eqnarray}
P({y}, {g})&=& P({y} \mid {g})  P({g})
\label{eq:2}
\end{eqnarray}
(where we define the right-hand side to equal 0 for any $g$ such that
$P(g) = 0$).  In the Bayes net of Eq.~\ref{eq:2}, the unconditioned
distribution identified with the parent node is $P(g)$, and the
conditioned distribution identified with the child node is $P(y \mid
g)$.  This Bayes net can be used to express Realist's reasoning.
Realist interprets the statements that `your choice occurs after $W$
has already made its prediction' and `when you have to make your
choice, you don't know what that prediction is' to mean that you can
choose any distribution $h({y})$ and then set $P({y} \mid {g})$ to
equal $h({y})$ (for all $g$ such that $P(g) \ne 0$). This is how
Realist interprets your having `free will'. (Note that this is a
different interpretation of `free will'' from the one made by
Fearful.) Under this interpretation, $W$ has no power to affect $P({y}
\mid {g})$.  Rather $W$ gets to set the parent node in the Bayes net,
$P(g)$.  For Realist, this is the distribution that you cannot
affect. (In contrast, in Fearful's reasoning, you set a
non-conditional distribution, and it is the conditional distribution
that you cannot affect.)

%Realist makes one
%presumption for each of
%the two distributions on the right-hand side of Eq.~\ref{eq:2}. (Just like Fearful makes two presumptions
%for the the terms on the right-hand-side of Eq.~\ref{eq:1}.) First, Realist presumes that your having
%`free will'' means that you can choose any distribution $h({y})$ and
%then set $P({y} \mid {g})$ to equal $h({y})$ (for all $g$ such that
%$P(g) \ne 0$).  (Note that this is a different interpretation of `free will'' from the one
%made by Fearful.) Under this presumption, $W$ has no power to affect $P({y} \mid {g})$. The
%second presumption is that your choice of $h$ does not affect $W$'s
%distribution $P(g)$, i.e., that $P(g)$ is `pre-fixed'' as far as you
%are concerned (as opposed to having $P(g \mid {y})$ be pre-fixed, as in
%the analysis below Eq.~\ref{eq:1}).

Realist's interpretation of Newcomb's paradox specifies what it is you
can set concerning $P({y}, g)$, and what is set by $W$. Just like
under Fearful's reasoning, under Realist's reasoning the `strategies' you
and $W$ choose do not directly specify the variables $g$ and
${y}$. Rather the strategies of you and $W$ specify two distributions
which, taken together, specify the full joint distribution. As before,
your task is to choose your strategy --- which now is $h({y})$ --- to
maximize your expected payoff under the associated $P({y}, {g})$.
Given Realist's presumption that the Bayes net of Eq.~\ref{eq:2}
underlies the game and that you get to set $h$, you should choose
$h({y}) = P({y} \mid {g}) = \delta_{{y}, AB}$, i.e., you should 
choose $AB$ with probability $1$. Doing this results in the expected
payoff $1,000 \; P(g=AB) \;+\; 1,001,000 \; P(g = B)$, which is your
maximum expected payoff no matter what the values of $P(g=AB)$ and
$P(g = B)$ are. This is the formal justification of Realist's
recommendation.

Note that in Fearful's interpretation, your strategy is choosing a single real number, 
while W chooses two real numbers. In contrast, in Realist's interpretation,
your strategy is still to choose a single real number, but now W also chooses
only a single real number. The different interpretations correspond to different 
games.{\footnote{In Realist's Bayes net, given the associated
restricted possible form of $P(y \mid g)$, there is no edge connecting the node for $g$ with the node for $y$. 
In other words, $P(y, g) = P(y)P(g)$. This means that $g$ and $y$ are `causally
independent', to use the language of causal nets~\cite{pear00}.  This causal relationship
is consistent with the single-stage game in
Table 1, in contrast to the causal relationship of the game played under Fearful's
interpretation, which requires a two-stage game.}}

\subsection{Combining the decompositions}

The original statement of Newcomb's question is somewhat informal. As originally stated
(and commonly analyzed), Newcomb's
question does not specify whether you are playing the game that Feaful
thinks is being played, or the game that Realist thinks is being
played. What happens if we try to formulate Newcomb's question
more formally? 

In light of extended game theory, there are three ways
we could imagine doing this. Let $\{S_y^F\}$ be your strategy set
in the game Fearful thinks is being played, and $\{S_y^R\}$ be your strategy set
in the game Realist thinks is being played. Similarly, let
$\{S_W^F\}$ be $W$'s strategy set
in the game Fearful thinks is being played, and $\{S_W^R\}$ be $W$'s strategy set
in the game Realist thinks is being played. 
\begin{enumerate}
\item The most natural way
to accommodate the informality of Newcomb's question is to define
an extended game where your strategy set is the combination of
what it is in Fearful's games and in Realist's game, i.e., where your
strategy set is $\{S_y^F\} \cup \{S_y^R\}$,
and similarly where $W$'s strategy set is $\{S_W^F\} \cup \{S_W^R\}$. 
\end{enumerate}

\noindent There are two additional ways it might be possible to merge
the games of Fearful and Realist:

\begin{enumerate}
\item[2.] Have your strategy set be $\{S_y^F\}$, while
$W$'s strategy set is  $\{S_W^R\}$. 
\item[3.] Have your strategy set be
$\{S_y^R\}$, while $W$'s strategy set is  $\{S_W^F\}$. 
\end{enumerate}

Are any of these extended games proper?
If not, then the only way of interpreting Newcomb's
scenario is as the game Fearful supposes or as the game Realist supposes. There
would be no way to combine those two games. 
%The resolution of Newcomb's question
%would then be immediate: Simply specify which of those two games are being played.

%Such a merging gives us an `extended game' of the sort considered
%in~\cite{wolp08b}, involving two Bayes net games. The Bayes nets of
%those two games are the two Bayes nets discussed in the preceding
%subsections. The strategy space of $W$ for the extended game is the
%set of all pairs of a conditional distribution $P(g
%\mid y )$ for the first (Fearful) Bayes net and a distribution $P(g)$ for the second
%(Realist) Bayes net. Your strategy space is the set of all pairs of a
%$P(y)$ for the first Bayes net, and a distribution $h(y)$ for the
%second Bayes net. 

We start by analyzing scenario [1]. 
If $P(g \mid {y})$ is set by W to be $\delta_{g,y}$
for all $y$ such that $P(y) \ne 0$, as under Fearful's
presumption, then your (Realist) choice of $h$ affects $P(g)$. In
fact, your choice of $h$ fully specifies $P(g)$.{\footnote{For
example, make Fearful's presumption be that $P(g \mid y) = \delta_{g,y}$
for all $y$ such that $P(y) \ne 0$. Given this, if you set $h({y}) =
\delta_{{y},AB}$, then $P(g) = \delta_{g,AB}$, and if you set $h({y})
= \delta_{{y},B}$, then $P(g) = \delta_{g,B}$.}} This contradicts
Realist's presumption that it is $W$'s strategy that sets $P(g)$,
independent of you. So scenario [1] is impossible; it is over-played, in the
terminology of Sec.~\ref{sec:improper}.

Now consider scenario [2]. In this scenario, any strategy $S_y$ specifies the
distribution $P(y)$ only, with $P(g \mid y)$ being arbitrary. In other words, such a strategy consists of all joint
distributions $P(y, g)$ with the specified marginal distribution $P(y)$. Similarly, any
strategy $S_W$ consists of all joint
distributions $P(y, g)$ with a specified marginal $P(g)$ and $P(y, g) = h(y) P(g)$ for some
$h$. For \emph{any} pair of such strategies $(S_y, S_W)$, there is a unique
joint distribution in $S_y \cap S_W$: the product distribution $P(y) P(g)$ where
$S_y$ sets $P(y)$ and $S_g$ sets $P(g)$. However this is just Realist's extended game;
it is not a new extended game.

We end by presenting two arguments that establish the impossibility of scenario [3].
First, we will show that if $W$ sets $P(g
\mid {y})$ for Fearful's Bayes net appropriately, then some of your
strategies $h(y)$ for Realist's Bayes net become impossible. In other
words, for those strategy choices by $W$ and by you, the joint
distribution over the game variables for Realists's Bayes net
contradicts the joint distribution over the game variables for
Fearful's Bayes net.  (This is true for almost any $P(g
\mid {y})$ that $W$ might choose, and in particular is true even if
$W$ does not predict perfectly.) 

To begin, pick any $\alpha \in (1/2, 1]$, and presume that $P(g
\mid {y})$ is set by $W$ to be $\alpha \delta_{g,y} + (1 - \alpha)(1 -
\delta_{g,y})$ (for all $y$ such that $P(y) \ne 0$). So for example,
under Fearful's interpretation, where $W$ uses a perfectly error-free
prediction algorithm, $\alpha = 1$. Given this presumption, the only
way that $P({y} \mid g)$ can be $g$-independent (for all $g$ such that
$P(g) \ne 0$) is if it is one of the two delta functions,
$\delta_{{y},AB}$ or $\delta_{{y},B}$. (See the appendix for a formal
proof.)  This contradicts Realist's interpretation of the game, under
which you can set $P({y} \mid g)$ to any $h({y})$ you
desire.{\footnote{Note that of the two $\delta$ functions you can
choose in this combined decomposition, it is better for you to choose
$h({y}) = \delta_{{y},B}$, resulting in a payoff of $1,000,000$. So
your optimal response to Newcomb's question for this variant is the
same as if you were Fearful.}}

Conversely, if you can set $P({y} \mid g)$ to be an arbitrary
$g$-independent distribution (as Realist presumes), then what you set
it to may affect $P(g \mid {y})$ (in violation of Fearful's
presumption that $P(g \mid {y})$ is set exclusively by W). In other
words, if your having `free will' means what it does to Realist, then
you have the power to change the prediction accuracy of $W$ (!). As an
example, if you set $P({y} = AB \mid g) = 3/4$ for all $g$'s such that
$P(g) \ne 0$, then $P(g \mid {y})$ cannot equal $\delta_{g,{y}}$.
So scenario [3] is also improper; ; it is over-played, in the
terminology of Sec.~\ref{sec:improper}.

Combining our analyses of scenarios [1] through [3] shows that there are
two ways to formalize Newcomb's scenario. You can be free
to set $P({y})$ however you want, with $P(g \mid {y})$ set by $W$, as Fearful presumes,
{\it{or}}, as Realist presumes, you can be free to set $P({y}
\mid {g})$ to whatever distribution $h({y})$ you want, with $P(g)$
set by $W$. It is not possible to play both games
simultaneously.
%{\footnote{In a variant of Newcomb's question, you
%first choose one of these two presumptions, and then set the associated
%distribution. If the pre-fixed distribution $P(g \mid {y})$ arising in
%the first presumption is $\delta_{g,{y}}$, then your optimal responses
%depend on the pre-fixed distribution $P(g)$ arising in the second
%presumption--- a distribution that is not specified in Newcomb's
%question. If $P(g)$ obeys $P(g = B) > .999$, then your optimal pair of
%choices are first to choose to set the distribution $P({y} \mid g)$ to
%some $h({y})$, and then to set $h({y}) =
%\delta_{{y}, AB}$. If this condition is not met, you should first choose to set $P({y})$, and
%then set it to $\delta_{{y}, AB}$.}}

The resolution of Newcomb's paradox is now immediate. 
%None of the three possible
%ways of combining the games of Fearful and Realist result in a proper game.
%So w
When formalizing Newcomb's scenario, we must pick either Realist's game or Fearful's. Once we pick the game,
the optimal choice of you is perfectly well-defined. The only reason that there 
appears to be a paradox is that the way Newcomb's question is phrased leads one
to think that somehow those two games are being combined --- and in fact combining
the games leads to mathematical impossibilities.

\subsection{Discussion}

It is important to emphasize that the impossibility of playing both
games simultaneously arises for almost any error rate $\alpha$ in the
$P(g \mid {y})$ chosen by $W$, i.e., no matter how accurately $W$
predicts. This means that the stipulation in Newcomb's paradox that
$W$ predicts perfectly is a red herring. (Interestingly, Newcomb
himself did not insist on such perfect prediction in his formulation
of the paradox, perhaps to avoid the time paradox problems.) The
crucial impossibility implicit in Newcomb's question is the idea that
at the same time you can arbitrarily specify `your' distribution
$P({y} \mid g)$ and W can arbitrarily specify `his'' distribution $P(g
\mid {y})$. In fact, neither of you two can set your distribution
without possibly affecting the other's distribution; you and $W$ are
inextricably coupled.

We also emphasize that vague concepts (e.g., `free will') or
controversial ones (e.g., `causality') are not relevant to the
resolution of Newcomb's paradox; it is not necessary to introduce
mathematizations of those concepts to resolve Newcomb's paradox.  The
only mathematics needed is standard probability theory, together with
the axioms of VonNeumann Morgenstern utility theory.

Another important contribution of our resolution is that it shows that
Newcomb's scenario does not establish a conflict between game theory's
dominance principle and its expected utility principle, as some have
suggested.  Indeed, as mentioned, we adopt the standard expected
utility axioms of game theory throughout our analysis. However nowhere
do we violate the dominance principle.

The key behind our avoiding a conflict between those two principles is
our taking care to specify what Bayes net underlies the game.
Realist's reasoning \emph{appears} to follow the dominance principle,
and Fearful's to follow the principle of maximizing expected
utility. (Hence the conflicting answers of Realist and Fearful appear
to illustrate a conflict between those two principles.) However
Realist is actually following the principle of maximizing expected
utility
\emph{for that Bayes net game for which Realist's answer is correct}. In
contrast, Realist's reasoning is an unjustified violation of that
principle for the Bayes net game in which Realist's answer is
incorrect. (In particular, Realist's answer does \emph{not} follow
from the dominance principle in that Bayes net game.) It is only by
being sloppy in specifying the underlying Bayes net game that it
appears that there is a conflict between the expected utility
principle and the dominance principle.

Note also that your utility function is irrelevant to the paradox and
its resolution. The fundamental contradiction built into Newcomb's
scenario involves the specification of the strategy sets of you and $W$.
The associated utility functions never appear.

Another note-worthy point is that no time variable occurs in our analysis of Newcomb's
paradox. Concretely, nothing in our analysis using Bayes nets and associated
games requires that $W$'s prediction
occur before your choice. This lack of a time variable in our analysis 
means we can assign times to the events in our analysis
any way we please, and the analysis still holds; the analysis is
time-reversal invariant. 

This invariance helps clarify the differences in the assumptions made
by Realist and Fearful.  The invariance means that both the formal
statement of the paradox and its resolution are unchanged if the
prediction occurs \emph{after} your choice rather than (as is
conventional) before your choice. In other words, $W$ could use data
accumulated up to a time \emph{after} you make your choice to
`retroactively predict' what your choice was.  In the extreme case,
the `prediction' algorithm could even directly observe your
choice. All of the mathematics introduced above concerning Bayes nets,
and possible contradictions still holds.

In particular, in this time-reversed version of Newcomb's scenario,
Fearful would be concerned that $W$ can \emph{observe} his choice with
high accuracy. (That's what it means to have $P(g \mid y)$ be a delta
function whose argument is set by the value of $y$.)  Formally, this
is exactly the same as the concern of Fearful in the conventional
Newcomb's scenario that $W$ can \emph{predict} his choice with high
accuracy.  

In contrast, in the time-reversed version of Newcomb's scenario,
Realist would believe that he can guarantee that his choice is
independent of what $W$ says he chooses.  (That's what it means to
have $P(y \mid g)$ equal some $h(y)$, independent of $g$.)  In
essence, he assumes that you can completely hide your choice from $W$,
so that $W$ can only guess randomly what choice you made. Formally,
this assumption is exactly the same as the `free will' belief of
Realist in the conventional Newcomb's scenario that you can force $W$'s
prediction to be independent of your choice. 

In this time-reversed version of Newcomb's scenario, the differences
between the assumptions of Realist and Fearful are far starker than in
the conventional form of Newcomb's scenario, as is the fact that those
assumptions are inconsistent. One could use this time-reversed version
to try to argue in favor of one set of assumptions or the other
(i.e., argue in favor of one Bayes net game or the other). Our intent
instead is to clarify that Newcomb's question does not specify which
Bayes net game is played, and therefore is not properly posed. As soon
as one or the other of the Bayes net games is specified, then
Newcomb's question has a unique, correct answer.

\section{Conclusion}

Newcomb's paradox has been so vexing that it has led some to resort to
non-Bayesian probability theory in their attempt to understand
it~\cite{giha78,huri78}, some to presume that payoff must somehow
depend on your beliefs as well as what's under the
boxes~\cite{gean97}, and has even even led some to claim that quantum
mechanics is crucial to understanding the paradox
\cite{pisl02b}. This is all in addition to work on the paradox based on now-discredited
formulations of causality~\cite{jaco93}.

Our analysis shows that the resolution of Newcomb's paradox is in fact
quite simple. Newcomb's paradox takes two incompatible interpretations
of a question, with two different answers, and makes it seem as though
they are the same interpretation. The lesson of Newcomb's paradox is
just the ancient verity that one must carefully define all one's
terms.

$ $

$ $

\noindent {\textbf{Acknowledgements}} We would like to thank Mark Wilber,
Charlie Strauss and an anonymous referee for helpful comments.

$ $

$ $

\newpage

\begin{eqnarray*}
 \begin{array}{ccc}
 &\multicolumn{2}{c}{} \\
  & {\underline{\textbf{Choose $AB$}}} & {\underline{\textbf{Choose $B$}}} \\
{}& & \\
{\textbf{Predict $AB$:}} & 1000 & 0 \\
 & & \\
{\textbf{Predict $B$:}} & 1,001,000 & 1,000,000
\label{ta:payoff}
\end{array}
\end{eqnarray*}
{\bf{Table 1:}} The payoff to you for the four combinations of your
    choice and $W$'s prediction.

\newpage

\noindent {\textbf{APPENDIX}}

$ $

\noindent In the text, it is claimed that if for some $\alpha \in (1/2, 1]$, $W$ can ensure that
$P({g} \mid {y}) =\alpha \delta_{g,y} + (1 - \alpha)(1 - \delta_{g,y})$ for all $y$ such that $P(y) \ne 0$,
and if in addition $P({y} \mid {g})$ is ${g}$-independent for
all $g$ such that $P(g) \ne 0$, then in fact $P(y \mid g)$ is one of the two delta
functions, $\delta_{{y},AB}$ or $\delta_{{y},B}$.  This can be seen
just by examining the general $2 \times 2$ table of values of $P(g, y)$
that is consistent with the first hypothesis, that
$P({g} \mid {y}) =\alpha \delta_{g,y} + (1 - \alpha)(1 - \delta_{g,y})$:

\begin{eqnarray*}
 \begin{array}{ccc}
 &\multicolumn{2}{c}{} \\
  & {\underline{\textbf{$y = AB$}}} & {\underline{\textbf{$y = B$}}} \\
{}& & \\
{\textbf{$g = AB$:}} & \alpha z_{AB} & (1 - \alpha)z_B \\
 & & \\
{\textbf{$g = B$:}} & (1-\alpha)z_{AB} & \alpha z_B
\label{ta:app_payoff}
\end{array}
\end{eqnarray*}
{\bf{Table 2:}} The probabilities of the four combinations of your
    choice $y$ and $W$'s prediction $g$, given that $P({g} \mid {y}) =\alpha \delta_{g,y} + (1 - \alpha)(1 - \delta_{g,y})$ for all $y$ with non-zero $P(y)$. Both $z_{AB}$ and $z_B$ are in $[0, 1]$, and normalization means that $z_{AB} + z_B = 1$.
 
$ $

If both $z_{AB} \ne 0$ and $z_B \ne 0$, then neither $P(g = AB)$ nor $P(g = B)$ equals zero. Under such a condition, the second hypothesis, that $P({y} \mid {g})$ is ${g}$-independent
for all $g$ such that $P(g) \ne 0$,
would mean that $P(y \mid g)$ is the same function of $y$ for both $g$'s. This in turn
means that 
\begin{eqnarray}
\frac{\alpha z_{AB}}{(1 - \alpha) z_{B}} &=& \frac{(1 - \alpha)z_{AB}}{\alpha z_{B}}
\end{eqnarray}
which is impossible given our bounds on $\alpha$. Accordingly, either $z_{AB}$ or $z_B$ equals zero.  \textbf{QED}

\newpage

\theendnotes

\newpage

%\bibliographystyle{naturemag}
%\bibliographystyle{amsplain}
% \bibliographystyle{unsrt}
%\bibliographystyle{apalike}
%\bibliographystyle{plainnat}
%bibliographystyle{harvard}
\bibliographystyle{klunamed}

\end{article}
\end{document}